\documentclass[nopreprint,twocolumn]{revtex4}
\usepackage{dcolumn}
\usepackage{graphicx}
\usepackage{amsmath}
\usepackage{amssymb}
\usepackage{citesort}
\usepackage{multirow}

\begin{document}
%
\title{Nucleation threshold and deactivation mechanisms of nanoscopic cavitation nuclei}

\author{Bram M. Borkent} \author{Stephan Gekle}
\author{Andrea Prosperetti} \altaffiliation[Also at: ]{Department of Mechanical Engineering, Johns Hopkins University, Baltimore, MD 21218, USA.} \author{Detlef~Lohse} \email{d.lohse@utwente.nl}
\affiliation{Physics of Fluids, Faculty of Science and Technology and J.M. Burgers Centre for Fluid Dynamics, Impact Institute, and MESA$^+$ Institute for Nanotechnology, University of Twente, P.O. Box 217, 7500 AE Enschede, The Netherlands.}

\date{\today}

\begin{abstract}
The acoustic nucleation threshold for bubbles trapped in cavities has theoretically been predicted within the crevice theory by Atchley \& Prosperetti [J. Acoust. Soc. Am. \textbf{86}, 1065-1084 (1989)]. Here, we determine this threshold experimentally, by applying a single pressure pulse to bubbles trapped in cylindrical nanoscopic pits ("artificial crevices") with radii down to 50\,nm. By decreasing the minimum pressure stepwise, we observe the threshold for which the bubbles start to nucleate. The experimental results are quantitatively in excellent agreement with the theoretical predictions of Atchley \& Prosperetti. In addition, we provide the mechanism which explains the deactivation of cavitation nuclei: gas diffusion together with an aspherical bubble collapse. Finally, we present superhydrophobic nuclei which cannot be deactivated, unless with a high-speed liquid jet directed into the pit.


\end{abstract}


\maketitle

\section{Introduction}
 Water can be ruptured at much smaller tensile stresses than theoretically expected~\cite{caupin06}. The reason for this discrepancy is the existence of small inhomogeneities in the liquid, which exist even when special care on the cleanliness of the water has been taken~\cite{herbert06}. The inhomogeneities, whatever their origin might be, have received the generic name "cavitation nuclei", while the bubble generation produced in this way is termed heterogeneous nucleation. Cavitation nuclei are generally long-lived and it is believed that they consist at least in part of a volume of gas~\cite{morch07}. This observation excludes the possibility of the inhomogeneities being free spherical gas bubbles, as these are unstable. To account for stable gaseous cavitation nuclei, two types are distinguished in the literature: bubbles stabilized by a skin (see Ref.~[\onlinecite{morch07}] and references therein) and bubbles trapped inside a surface defect ("crevice model")\cite{harvey44,strasberg59,apfel70,crum79,atchley89,morch00}. 

The principle of the crevice model dates back to 1944~\cite{harvey44} and has found extensive qualitative experimental evidence over the years. Greenspan and Tschiegg~\cite{greenspan67}, for example, reported that removing particles larger than 0.2$\mu$m in diameter increased the tensile strength of water to about 200 bar (see also Refs.~[\onlinecite{briggs50, herbert06}]). Others found that the addition of suspended particles lowers the nucleation threshold~\cite{roy89,madanshetty91a,madanshetty91b,madanshetty95,deng96,marschall03}, while pre-experimental pressurization of water increases the nucleation threshold\cite{strasberg59,borkent07a}. Although these findings are in line with the general idea of the crevice model, none of the experiments could quantitatively verify the theoretical crevice model as developed by Atchley \& Prosperetti~\cite{atchley89} in 1989. One of the reasons is that their predictions are valid for a single cavity of a well-defined shape, while in practice the liquid usually contains a wide variety of nuclei of different sizes and shapes. Even in ultrapure water with a controlled number of microparticles, the sizes of the nuclei present on the microparticles can exhibit size variations yielding a wide range of thresholds~\cite{borkent07a}.

A step forward was achieved by Bremond~\emph{et al.} who were able to create monodisperse cavitation nuclei by trapping gas inside cylindrical holes of well-defined shape etched in silicon surfaces using standard lithography techniques~\cite{bremond05,bremond06a,bremond06b}. Not only was the position and size of the nuclei perfectly controlled, but the nucleation event itself was also highly reproducible, so that it could be followed in time with stroboscopic methods without the need of expensive high-speed cameras.

Both conditions, the reproducibility of the experiment and the monodispersity of the nuclei present at fixed positions, are important ingredients of this paper. Here, we have downscaled the micropits of Bremond \emph{et al.} two orders of magnitude, so that it becomes possible to experimentally verify the theoretical predictions made in the framework of the crevice model. This is the first aim of this paper. Secondly, we explore the mechanisms leading to the deactivation of nuclei after a single nucleation event. In addition, we show that superhydrophobic cavitation nuclei can nucleate hundreds of times, without being deactivated. Our observations and interpretations have implications for an increased understanding of the behavior of cavitation nuclei down to length scales of a few tens of nm ("surface nanobubbles")~\cite{borkent07b}.

\section{Brief theoretical description}
\begin{figure}
\centering
        \includegraphics[width=50mm]{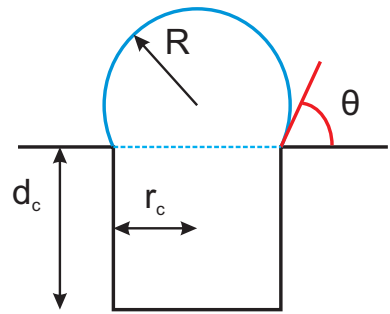}
        \caption{Cylindrical cavity with its dimensions. The initial gas-liquid interface is flat (dashed line), while the expanding bubble has a radius of curvature $R$ and contact angle $\theta$ with the flat surface.}
        \label{fig_nanopit}
\end{figure}

A comprehensive description and development of the crevice model can be found in the paper of Atchley \& Prosperetti~\cite{atchley89} with extensions to any axisymmetric geometry, including cylindrical cavities, given by Chappell \& Payne~\cite{chappell07}. In both papers, the authors predict the various nucleation thresholds as function of crevice shape, gas tension, and receding contact angle. Atchley \& Prosperetti argue that a true nucleation event must be the result of the loss of mechanical stability of the nucleus, and calculated the threshold for two situations: the first threshold denotes the pressure at which the nucleus \emph{inside }a crevice begins its unstable growth, and is the lowest pressure value belonging to a bubble reaching either the critical radius of curvature $R_c$, or the receding radius of curvature $R_R$, i.e. the radius of curvature at which the receding contact angle $\theta_R$ is reached. The second threshold is equivalent in definition, but holds for the bubble growing \emph{outside }of the crevice mouth. The lower value of the first and second nucleation threshold is the one for which the bubble grows explosively out of the cavity. For the case of a bubble trapped in a cavity with volume $V_c$ with its interface at the crevice mouth, the (second) nucleation threshold is given by

\begin{equation}\label{eq1}
p_L+\frac{2\sigma}{R}=p_v + \frac{V_0 p_{g,0}}{V_c+(\pi / 3)g(\theta)R^3},
\end{equation}

\noindent with $p_L$ the liquid pressure, $p_v$ the vapor pressure, $V_0$ the initial volume of the gas, $p_{g,0}$ the initial gas pressure in the bubble, $\sigma$ the liquid-gas surface tension, and $(\pi/3)g(\theta)R^3$ the volume of the spherical-cap-shaped bubble with radius $R$ as it expands above the cavity~\cite{atchley89} (see Fig.~\ref{fig_nanopit}). Here, $g(\theta)=\left(2+(2+\sin^2\theta)\cos\theta\right)$ is a geometric function depending on the contact angle $\theta$. The right-hand side of Eq.~\ref{eq1} represents the expanding forces caused by the vapor and gas pressure, $p_v + p_g$, respectively, while the left-hand side represents the collapsing forces due to the liquid pressure and surface tension, respectively. Eq.~\ref{eq1} implies that, for nucleation to occur, the expanding forces should exceed the collapsing forces (condition 1). Secondly, this condition should persist for increasing $R$, i.e. $\mathrm{d}(p_g + p_v)/\mathrm{d}R > \mathrm{d}(p_L + 2\sigma /R)/\mathrm{d}R $ (condition 2).

Let us now consider the case - which we will examine experimentally - of a cylindrical cavity with radius $r_c$ and depth $d_c$, under the assumption that the initial gas-liquid interface at the crevice mouth is approximately flat (i.e. we assume a negligible effect of the hydrostatic pressure and gas-saturated water), so that we can write $V_0 = V_c=\pi r_c^2 d_c$. For $\theta_R\geq\pi/2$ the nucleation threshold is the pressure needed to pull the bubble beyond its minimum radius while it expands from the cavity, i.e. $R_{\mathrm{min}}=R_R=r_c/\sin\theta_R$. Now, Eq.~\ref{eq1} can be rewritten as

\begin{equation}\label{eq2}
p_L=p_v + \frac{3 p_{g,0}}{3 + (r_c/d_c)g(\theta_R)/\sin^3{\theta_R}}-\frac{2\sigma\sin{\theta_R}}{r_c}.
\end{equation}

\begin{figure}
\centering
        \includegraphics[width=70mm]{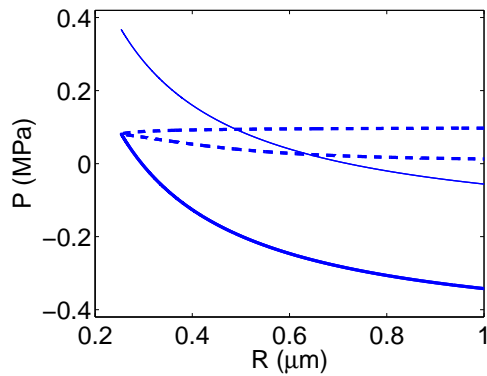}
        \caption{Graph of the expanding forces (right-hand side of Eq.~\ref{eq1}, dashed line) and the collapsing forces (left-hand side of Eq.~\ref{eq1}, solid lines) for a bubble expanding from a cylindrical pit with $r_c=250\,\rm{nm}$, $d_c=500\,\rm{nm}$, $p_{g,0}=10^5\,\mathrm{Pa}-p_v$, $p_v=73 \cdot 10^2\,\mathrm{Pa}$ and $\theta_R=100^{\rm{o}}$. The expanding forces (dashed line) show two branches corresponding to the possible solutions of $R$. The upper branch reflects the initially flat bubble during its first expansion phase: $R$ decreases from $R=R_{\infty}$ to $R=R_R$; the lower branch shows the solutions for the bubble expanding during its second phase: $R$ increases from $R=R_R$ to larger sizes. The collapsing forces are shown for two cases: 1) $p_L=-0.20\,\rm{MPa}$ (thin solid line) and 2) $p_L=-0.486\,\rm{MPa}$ (thick solid line). For case 1, the expanding forces dominate over the collapsing forces and the bubble will expand, until at $R=0.49\mu \rm{m}$ a stable equilibrium is reached. No nucleation will occur in this case. In case 2 the expanding forces are larger than the collapsing forces for all possible solutions of $R$, and as a result the bubble will grow explosively.}
        \label{fig_threshold}
\end{figure}

This prediction will be verified experimentally in this paper. At the point $R=R_{\mathrm{min}}=R_R$ the bubble experiences the maximum collapsing force. Any further reduction of the liquid pressure will make the bubble expand, thus reducing the surface tension pressure, and the bubble will grow explosively with the contact angle fixed at $\theta_R$. Equations~\ref{eq1} and~\ref{eq2} are illustrated in Fig.~\ref{fig_threshold}, for an air bubble trapped in a cylindrical pit with dimensions $r_c=250\,\rm{nm}$ and $d_c=500\,\rm{nm}$ and with $p_{g,0}=10^5\,\mathrm{Pa}-p_v$, $p_v=73 \cdot 10^2\,\mathrm{Pa}$ and $\theta_R=100^{\rm{o}}$. The graph shows the expanding and the collapsing forces for two cases: 1) $p_L=-0.20\,\rm{MPa}$ (Blake threshold~\cite{brennen95,leighton94} for a free bubble with $R_0=r_c$) and 2) $p_L=-0.486\,\rm{MPa}$ (prediction of Eq.~\ref{eq2}). From this plot it is readily seen that the expanding pressure in the second case is always larger than the collapsing pressure, and that $\mathrm{d}(p_g+p_v)/\mathrm{d}R  > \mathrm{d}(p_L+2\sigma/R)/\mathrm{d}R $.

For $\theta_R<\pi/2$ the nucleation threshold is much more complicated to calculate. At $R=R_{\mathrm{min}}=r_c$ the collapsing force due to surface tension is indeed maximum, but now $\mathrm{d} (p_g+p_v)/\mathrm{d}R < \mathrm{d}(p_L+2\sigma/R)/\mathrm{d}R $ and therefore $p_L(R_\mathrm{min})$ cannot be the nucleation threshold. Also $p_L(R_R)$ is not the correct threshold, as this gives a stable equilibrium in the upper branch of the compressive force curve. Instead, the threshold needs to be found by numerically solving $\mathrm{d}(p_g+p_v)/\mathrm{d}R  = \mathrm{d} (p_L+2\sigma/R)/\mathrm{d}R$, which is an implicit equation as the contact angle depends on the radius $R$, through $\sin\theta=r_c/R$, as long as the receding contact angle has not yet been reached.

In the prediction of $p_L$ the gas term is significant for cavities down to a few hundred nm in radius and will therefore be taken into account in the present analysis.


\section{Materials \& methods}
\subsection{Experimental setup}
\begin{figure}
\centering
        \includegraphics[width=70mm]{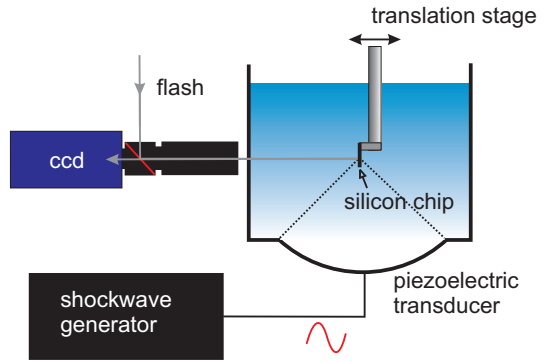}
        \caption{Sketch of the experimental setup.}
        \label{fig_setup}
\end{figure}

\begin{figure}
\centering
        \includegraphics[width=70mm]{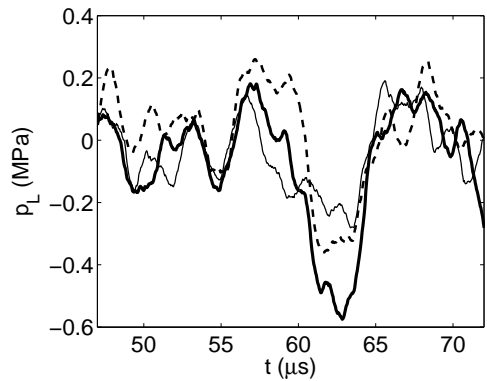}
        \caption{Three pressure signals with increasing strength recorded at the chip surface 25\,mm out of focus, corresponding to the experiment with sample B ($r_c=246\,$nm). Each line is the mean of 25 recordings. From these signals the minimum pressure can be extracted: -0.24\,MPa (thin solid line), -0.35\,MPa (dashed line) and -0.54\,MPa (thick solid line)}
        \label{fig_P_246}
\end{figure}

The experimental setup to investigate the nucleation behavior of bubbles trapped in well-defined cavities is similar to that used by Bremond~\emph{et al.}~\cite{bremond05,bremond06a,bremond06b} and sketched in Fig.~\ref{fig_setup}. Cavitation is induced by a focused shock wave generator (Piezoson 100, Richard Wolf GmbH) consisting of piezoelectric elements mounted on a spherical cap at the bottom of the liquid bath, which is filled with 1 liter of water (Milli-Q Synthesis A10, Millipore). The cavitation activity is recorded optically with a CCD camera (Flowmaster, LaVision) through a long-distance microscope (Model K2, Infinity). Illumination is provided by a flash lamp in reflection mode. The liquid pressure $p_L$ is obtained with the help of a calibrated glass fiber hydrophone (FOPH 500, RP Acoustics). The pressure is derived by measuring the reflected intensity of the laser beam at the fiber tip, which depends on the density of the water as affected by the local pressure~\cite{staudenraus93}. At the acoustic focus the pressure signal is typically characterized by a pressure peak (duration $\sim1\,\mu s$) followed by a negative pressure phase ($\sim 5 \mu s$). The intensity of the pressure pulse can be varied in twenty discrete steps. Since the smallest possible pressure decrease at the acoustic focus min$(p_L)=-3.2\,$MPa is already too large for our purpose, the samples are translated horizontally (away from the acoustic focus) along the line of sight, until the pressure signal is sufficiently weak that nucleation does not occur at the smallest pressure drop, but only at larger pressure decreases. The corresponding pressure signals are recorded using a low pass filter and averaged over 25 recordings to reduce the noise. A typical recording of the pressure signal obtained 25\,mm out of focus is shown in Fig.~\ref{fig_P_246} and corresponds to the experiment with the sample containing pits of 246\,nm in radius.

\subsection{Samples with nanopits}
\begin{figure}
\centering
        \includegraphics[width=85mm]{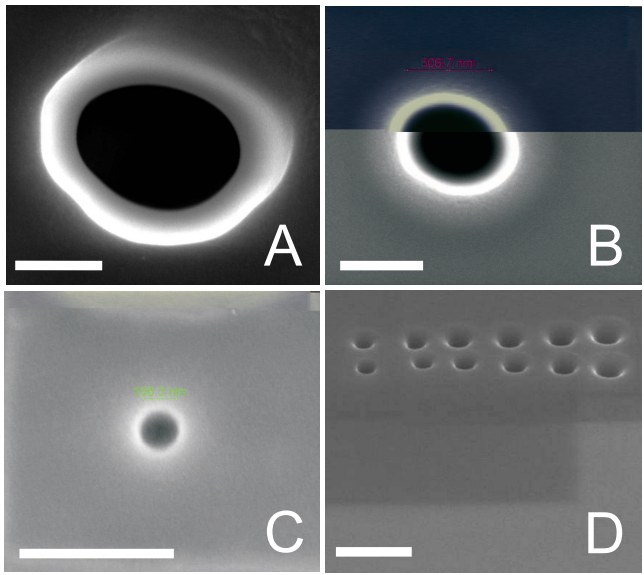}
        \caption{SEM images of samples A-D, respectively. The bars in each picture denote 500\,nm. The ellipsoidal shapes for sample A and B are due to drift.}
        \label{fig_nanopits_sem}
\end{figure}

The substrates of interest are silicon pieces of $5\times5\,\mathrm{mm}^2$ diced from a Si(100) wafer. The nanoscopic cylindrical pits are directly etched into the substrate by a focused ion beam (FIB) in a $6\times6$ square pattern, with $200\,\mu \mathrm{m}$ distance between the pits. The resulting $1\times1\,\mathrm{mm}^2$ pattern is located at the center of the chip. In order to facilitate detailed imaging by atomic force microscopy (AFM) and scanning electron microscopy (SEM) identical pits are etched the chip corner. Four samples (A-D) were studied with the following dimensions (determined with SEM) of the nanopits: A) $r_c= 495$\,nm; B) $r_c= 246$\,nm; C) $r_c= 53$\,nm; D) $r_c=50-60$\,nm. In sample A-C the pattern consisted of uniformly sized pits, with depth $d_c=2r_c$, while in sample D each column of pits had different depths (75, 100, 200, 300, 500, and 1000\,nm), which influenced the radial pit size per column by a few nm (50, 50, 50, 55, 57, and 60\,nm resp.); see Fig.~\ref{fig_nanopits_sem} for the corresponding SEM pictures. After production the samples were cleaned ultrasonically in ethanol (15 minutes), followed by an oxygen plasma (5 min.), a chemical cleaning step using a fresh (5:1) Piranha mixture (30 min.) and again an ultrasonic bath in ethanol (15 min.). This yielded clean and completely wetting substrates, which were characterized by a smoothly dewetting contact line (if not, the whole process was repeated). Subsequently, the samples were hydrophobized with 1-$H$,1-$H$,2-$H$,2$H$-perfluorodecyltrichlorosilane following Ref.~[\onlinecite{mayer00}]. The advancing and receding contact angles on the surface were $\theta_a=124^{\rm{o}}$ and $\theta_r=100^{\rm{o}}$. After immersion in water it was confirmed with AFM in tapping mode that a horizontal gas-liquid meniscus was present at the mouth of the pits.

\section{Results}
\subsection{Nucleation from gas-filled nanopits}
\begin{figure}
\centering
        \includegraphics[width=85mm]{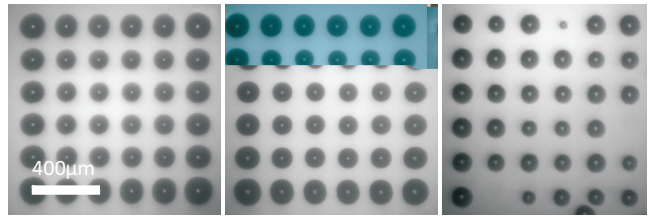}
        \caption{Cavitation bubbles nucleated from cylindrical pits with radius $r_c=495$\,nm (left), $r_c=246$\,nm (middle) and $r_c=53$\,nm (right), and depth $d_c=2r_c$ for a pressure pulse with $p_m=-3.2\,\mathrm{MPa}$.}
        \label{fig_cav_focus}
\end{figure}

To see whether it is possible to nucleate bubbles from nanopits as small as 53\,nm in radius, samples A-C were immersed in the liquid bath allowing air to be entrapped in the pits. In successive experiments the samples were put at the acoustic focus of the shock wave generator and subjected to a pressure pulse with $p_m=\mathrm{min}(p_L)=-3.2\,\mathrm{MPa}$. This value is sufficiently below the nucleation thresholds of the three samples, i.e. -0.23\,MPa, -0.48\,MPa, and -2.59\,MPa respectively (see Eq.~\ref{eq2}), to expect nucleation of bubbles from the nanopits. The camera and flash were triggered a few $\mu \mathrm{s}$ after passage of the negative part of the shock wave to capture the expanding bubbles at maximum sizes. The result is depicted in Fig.~\ref{fig_cav_focus} for the samples A-C, respectively. Sample A and B showed almost perfect bubble patterns, with each bubble corresponding to the position of the nanoscopic cavitation nucleus. In each experiment, the cavitation nuclei had to be 're-activated' (filled with air again), since it was not possible to nucleate bubbles a second time without taking the sample out of the water first~\cite{bremond06a}. With sample C a maximum amount of 34 bubbles could be nucleated in the first experiment, implying that it is indeed possible to nucleate bubbles from such small cavities. While sample A and B showed perfect reproducibility, the number of bubbles nucleating from sample C declined dramatically in later experiments, even when the negative pressure amplitude was increased to -7\,MPa. Presumably, small contaminant molecules had decreased the contact angle locally, leading to completely wetted nanopits. To test this possibility, the old hydrophobic coating was stripped off with an oxygen plasma and the sample was hydrophobized again through the cleaning and coating steps described before. This process indeed re-activated part of the nuclei ($\sim 80 \%$ of the pits) though the number of bubbles declined again in successive experiments.

\subsection{Determination of the experimental nucleation threshold}
The experimental nucleation threshold of the nanosized cavities can be obtained by moving the samples to a position in the liquid bath where the pressure drop $p_m$ is sufficiently small that no nucleation occurs. By lowering $p_m$ stepwise the cavities will nucleate at a certain negative pressure amplitude which is the experimental nucleation threshold. To observe the bubbles optically, it is not sufficient for the negative pressure to be low enough, but it should also last long enough in time. A lower limit to the time $\Delta t$ the bubble needs to grow to visible size is estimated, by first estimating the critical size $R_{c,o}$ to be optically observable. We take $R_{c,o}\sim 3$ image pixels = 3 pix $\times$ 2.9 $\mu \mathrm{m}$ /pix = 8.7 $\mu \mathrm{m}$. Now, using~\cite{bremond06b,plesset77}
$\dot{R}=\left(\frac{2}{3}\frac{p_v-p_m}{\rho}\right)^{1/2}$
with $\rho$ the liquid density and $\dot{R}$ the bubble wall velocity, it follows that $\Delta t=R_{c,o}/\dot{R} = 0.9\,\mu s$ (sample A), $0.6\,\mu s$ (sample B) and $\sim0.3\,\mu s$ (sample C and D). The minimum pressure level which lasts $\Delta t$ is the negative pressure amplitude $p_m$ of interest. Note that the difference with the absolute minimum pressure level is in most cases only a few percent.

\begin{figure}
\centering
        \includegraphics[width=80mm]{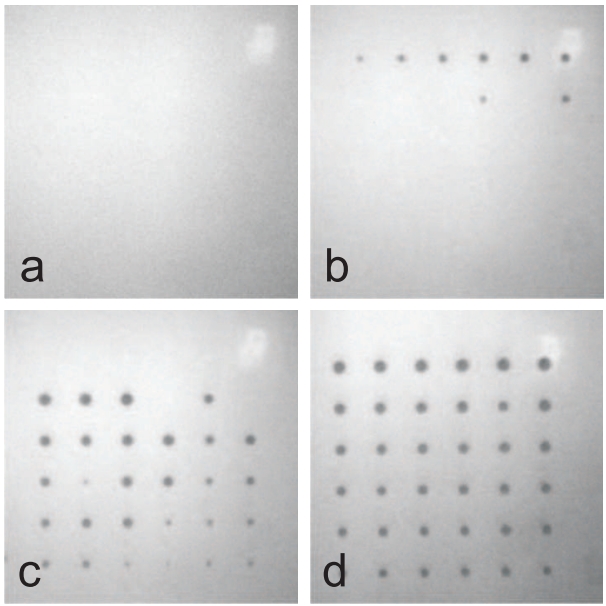}
        \caption{Cavitation bubbles emerging from $6\times6$ cylindrical pits with $r_c=246$\,nm (sample B), for three successively applied pressure pulses: a) $p_m=-0.24$\,MPa; b) $p_m=-0.35$\,MPa; c) $p_m=-0.54$\,MPa. The full pattern develops when $p_m=-0.54$\,MPa is applied without the other two preceding pulses (d).}
        \label{fig_exp_246}
\end{figure}

Let us first consider the case of sample B ($r_c=246\,$nm, $d_c=2r_c$). The sample was put 25\,mm out of the acoustic focus, and three pressure pulses with increasing strength (Fig.~\ref{fig_P_246}) were applied successively, without taking the sample out of the water. A typical result is shown in Fig.~\ref{fig_exp_246}. With the first pulse ($p_m=-0.24\,$MPa) no cavitation bubbles could be observed (Fig.~\ref{fig_exp_246}a). The second pressure pulse ($p_m=-0.35\,$MPa), resulted in a few nucleated bubbles, but the majority of the nuclei in the pattern still did not cavitate (Fig.~\ref{fig_exp_246}b). The third pressure pulse, with $p_m=-0.54\,$MPa, was able to nucleate all remaining nuclei (Fig.~\ref{fig_exp_246}c). Note that the nuclei which nucleated with the second pulse could not be nucleated with the stronger third pulse, implying that the nuclei can be used only once. Nuclei which were not nucleated during the second pulse, however, survived and were nucleated with the stronger third pulse. The experiment was repeated eight times with reproducible results (Table~\ref{tab:results}): on average 0, 10, and 26 bubbles were counted for the three applied pressure pulses, respectively. When the third pulse ($p_m=-0.54\,$MPa) was applied without the other two preceding ones, the full pattern became visible (Fig.~\ref{fig_exp_246}d). Since the majority of the pits nucleated at $p_m=-0.54\,$MPa it is reasonable to assume that this pressure amplitude is above the experimental nucleation threshold for sample B, while the weaker pulse ($p_m=-0.35\,$MPa) is (just) below the experimental nucleation threshold.

\begin{table*}[!btp]
\centering
\begin{tabular}{|c| p{1cm} p{1cm} p{1cm} p{1cm} p{1cm} p{1cm} p{1cm} p{1cm} |c|}
\hline \multirow{2}{*}{$p_m$ (MPa)} & \multicolumn{8}{c}{Amount of bubbles in exp. 1-8} \vline & \multirow{2}{*}{mean}\\

 & 1 & 2 & 3 & 4 & 5 & 6 & 7 & 8 & \\

\hline
\hline -0.24 & 0  & 0  & 0  & 0  & 0  &  0 & 0  & 0  & 0 \\
\hline -0.35 & 11 & 9  & 8  & 15 & 10 & 8  & 8  & 7  & 10\\
\hline -0.54 & 23 & 27 & 28 & 19 & 25 & 27 & 27 & 29 & 26\\
\hline

\end{tabular}

\caption{\label{tab:results}Results of eight experiments with sample B ($r_c=246\,$nm, $d_c=2r_c$). In each experiment the minimum pressure $p_m$ is decreased in three successive steps. The majority of the pits nucleates at $p_m=-0.54$\,MPa.
}
\end{table*}

\begin{figure}
\centering
        \includegraphics[width=80mm]{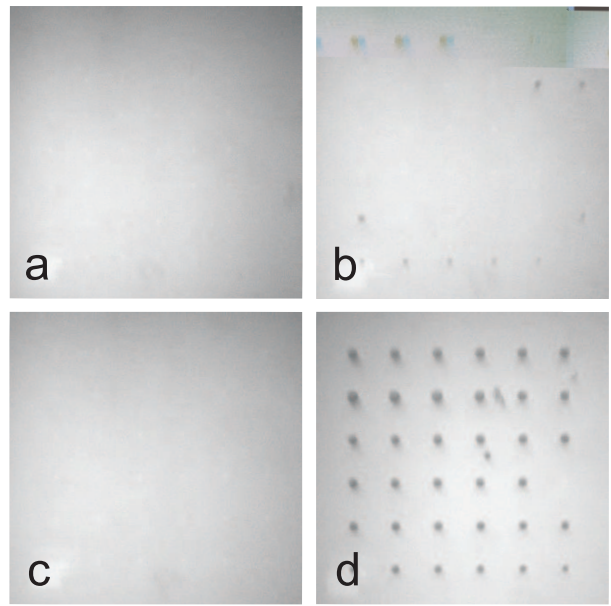}
        \caption{Cavitation bubbles emerging from $6\times6$ cylindrical pits with $r_c=495$\,nm (sample A), for three successively applied pressure pulses: a) $p_m=-0.20$\,MPa; b) $p_m=-0.23$\,MPa; c) $p_m=-0.34$\,MPa. The full pattern develops when $p_m=-0.34$\,MPa is applied without the other two preceding pulses (d).}
        \label{fig_exp_495}
\end{figure}

A similar experiment was carried out with sample A ($r_c=495\,$nm, $d_c=2r_c$, Fig.~\ref{fig_exp_495}). Again the sample was subjected to three successive pressure signals of decreasing negative pressure without being taken out of the water. For the lowest pressure amplitude ($p_m=-0.20\,$MPa) no cavitation bubbles could be detected optically (Fig.~\ref{fig_exp_495}a). A larger amplitude of $p_m=-0.23\,$MPa yielded 14 bubbles of different sizes, with some of them barely visible (Fig.~\ref{fig_exp_495}b), while a further reduction of the liquid pressure ($p_m=-0.34\,$MPa) resulted in no visible bubbles at all (Fig.~\ref{fig_exp_495}c). What happened with the remaining $36-14=22$ pits? As the lowest negative pressure was not able to nucleate them, they must already have been nucleated during the first two pulses, i.e. the nucleation took place below optical resolution. This is possible as the resolution of our optical detection is limited and the pressure pulse in this case is relatively weak (i.e. $\dot{R}$ is small). Therefore, in contrast with case B, we are not able to measure the pressure for which nucleation does \emph{not }take place. When the third pulse ($p_m=-0.34\,$MPa) was applied without the other two preceding pulses, the full pattern became visible (Fig.~\ref{fig_exp_495}d). Hence, this is the pressure level for which we are sure that full nucleation takes place.

\begin{figure}
\centering
        \includegraphics[width=67.5mm]{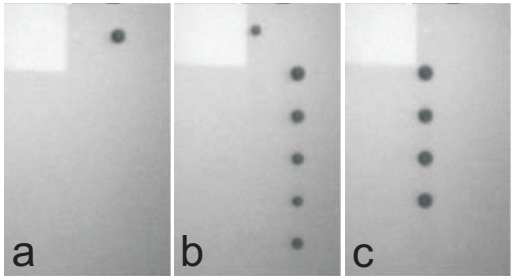}
        \caption{Cavitation bubbles emerging from $2\times6$ cylindrical pits (sample D) with $r_c=57$\,nm, $d_c=500$\,nm (left column) and $r_c=60$\,nm, $d_c=1000$\,nm (right column), for three successively applied pressure pulses: a) $p_m=-2.3$\,MPa; b) $p_m=-2.6$\,MPa; c) $p_m=-3.0$\,MPa.}
        \label{fig_exp_53}
\end{figure}

Finally, sample D was studied, which consists of nanopits with $50\,\mathrm{nm}\leq r_c \leq60\,$nm and varying depths. Just like sample C, the shallow pits on sample D nucleated only a few times, and could not be nucleated in later experiments. Fortunately, two columns with the deepest pits ($d_c=500\,\rm{nm}$ \& $r_c=57\,$nm, and $d_c=$1000\,nm \& $r_c=60\,$nm) could be nucleated repeatedly, and the nucleation threshold could be measured for these pits. The experiment was very similar to the ones described before, but now the sample was 12\,mm away from the acoustic focus. A typical experimental result is depicted in Fig.~\ref{fig_exp_53}. First, a pressure pulse with $p_m=-2.3\,$MPa was applied and 1 pit from the right column ($r_c=60$\,nm) was nucleated. A stronger second pulse ($p_m=-2.6\,$MPa) was able to nucleate the remaining 5 pits from this column, though other pits in the sample did not nucleate, as they were smaller. Reducing the negative pressure further ($p_m=-3.0\,$MPa) resulted in the nucleation of the left column of pits with $r_c=57\,$nm. Hence, a small variation in pit sizes of just a few nm is reflected in a different nucleation threshold. It is also observed that the pits did not nucleate a second time, despite their huge aspect ratios.

 \subsection{Comparison with theoretical prediction}
\begin{figure}
\centering
        \includegraphics[width=80mm]{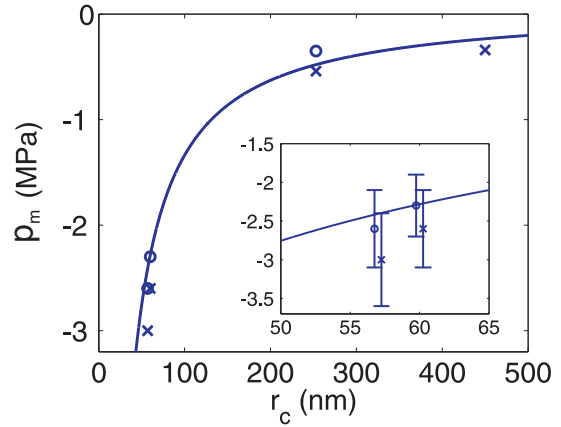}
        \caption{Nucleation threshold as function of the pit radius for both theory (line) and experiment (symbols, crosses: nucleation, circles: no nucleation). The inset shows a zoom in with errorbars. For visibility overlapping points are shifted $\pm$0.25\,nm with respect to each other.}
        \label{fig_thebigresult}
\end{figure}

How do the experimental results compare with theoretical predictions? In Fig.~\ref{fig_thebigresult} the theoretical nucleation threshold (line), based on Eq.~\ref{eq2}, is plotted as a function of the pit radius $r_c$ together with the experimental results (symbols). We used $d_c=2r_c$ as is the case in sample A and B. Note that for the pits present in sample D the gas pressure term is negligible: the difference between $d_c=2r_c$ and $d_c=20r_c$ changes the theoretical prediction for pits of $r_c=50\,$nm $<1$\%. The experimental data points at which full nucleation was detected for samples A, B and D are depicted by crosses, while the experimental pressures where full nucleation (just) did not happen are marked with circles. The inset shows the experimental results for sample D including typical error bars depicting the standard deviation of the pressure recordings.

We observe a striking quantitative agreement between theory and experiment for all samples. Pressure amplitudes for which nucleation was first detected are below the line marking the nucleation threshold. Pressure amplitudes for which nucleation did not occur are either above this line, i.e. in the regime where nucleation is not expected, or the line is within experimental errorbars. For sample D it was observed that the nucleation threshold strongly depends on the size of the pits: pit radii just a few nm smaller resulted in a significantly lower nucleation threshold, in agreement with the steep slope of $p_m (r_c)$ around these values.

\subsection{Deactivation of cavitation nuclei}
 It is well known that artificial nucleation sites in boiling continue to be active for a long time, emitting many bubbles~\cite{dhir98}. Similarly, the microscopic wall cracks and scratches in a glass full of beer or champagne are seen to emit bubbles for a very long time~\cite{liger08}. Even in cavitation studies on bare hydrophobic substrates, bubbles trapped in localized defects could be nucleated more than a hundred times~\cite{bremond05}. Thus, there is something special in the deactivation of nuclei observed here which makes this situation different from the others. What is the physical mechanism responsible for the deactivation of the nuclei?

  The main mechanism responsible for the deactivation of nuclei is encountered during the collapse phase of the bubble, where a wall-directed jet is formed. In the case of a single bubble (or weak interaction among bubbles) the jet momentum is expected to be directed mostly normal to the wall~\cite{benjamin66,plesset66}. When bubble-bubble interaction is non-negligible, on the other hand, the jet momentum may be deflected away from the normal~\cite{bremond06b}.

\begin{figure}
\begin{center}
\includegraphics[width=0.45\columnwidth]{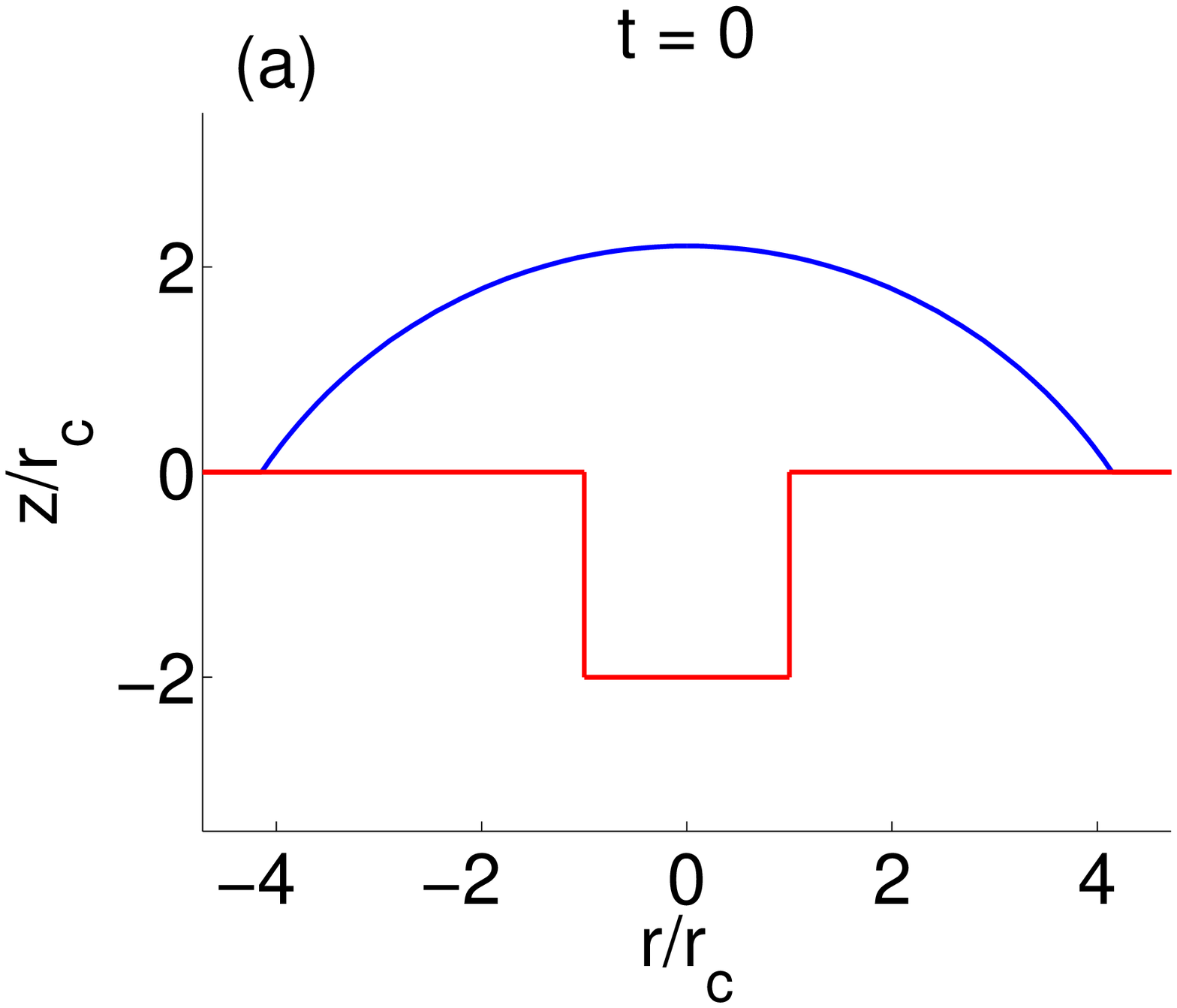}
\includegraphics[width=0.45\columnwidth]{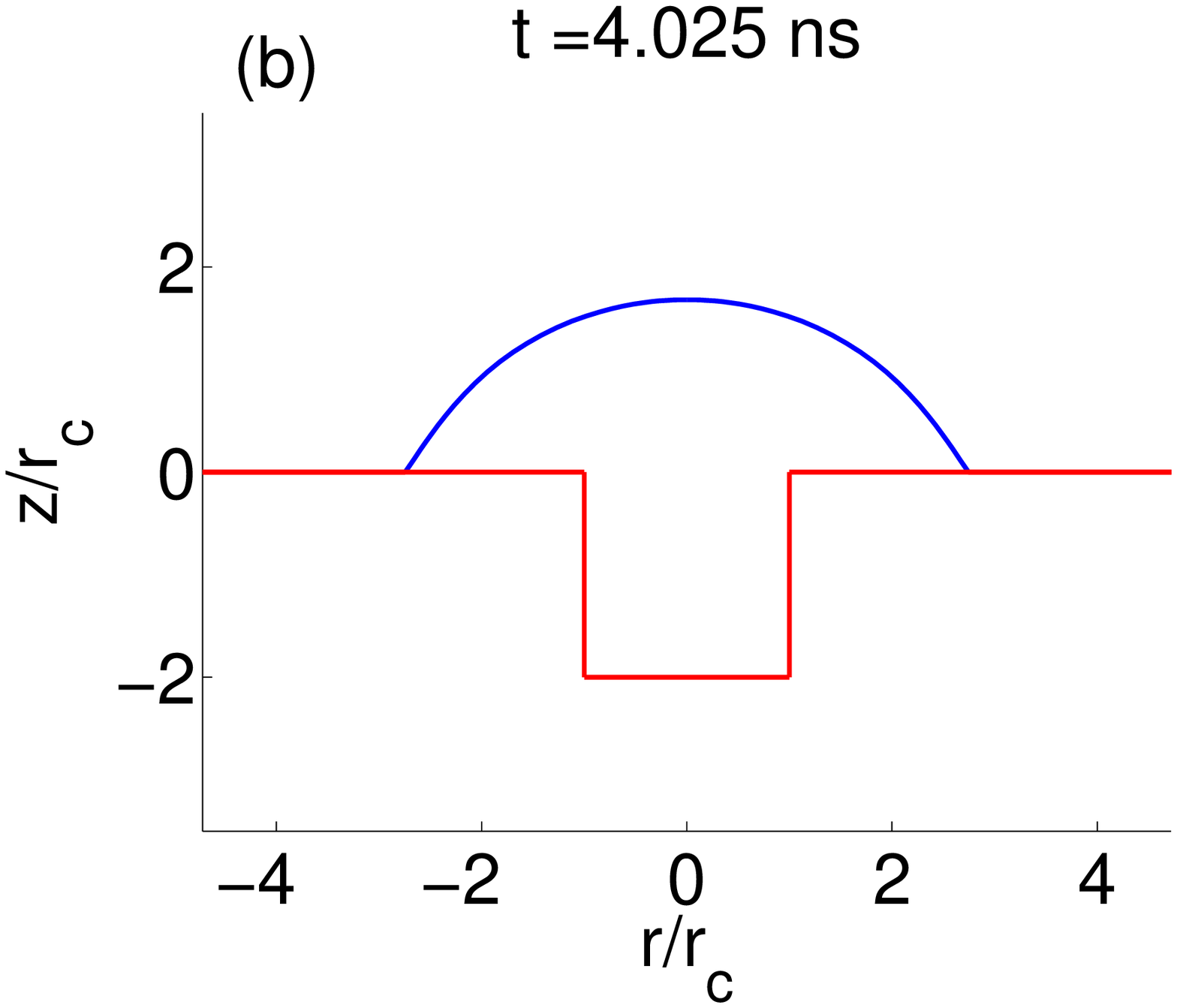}
\includegraphics[width=0.45\columnwidth]{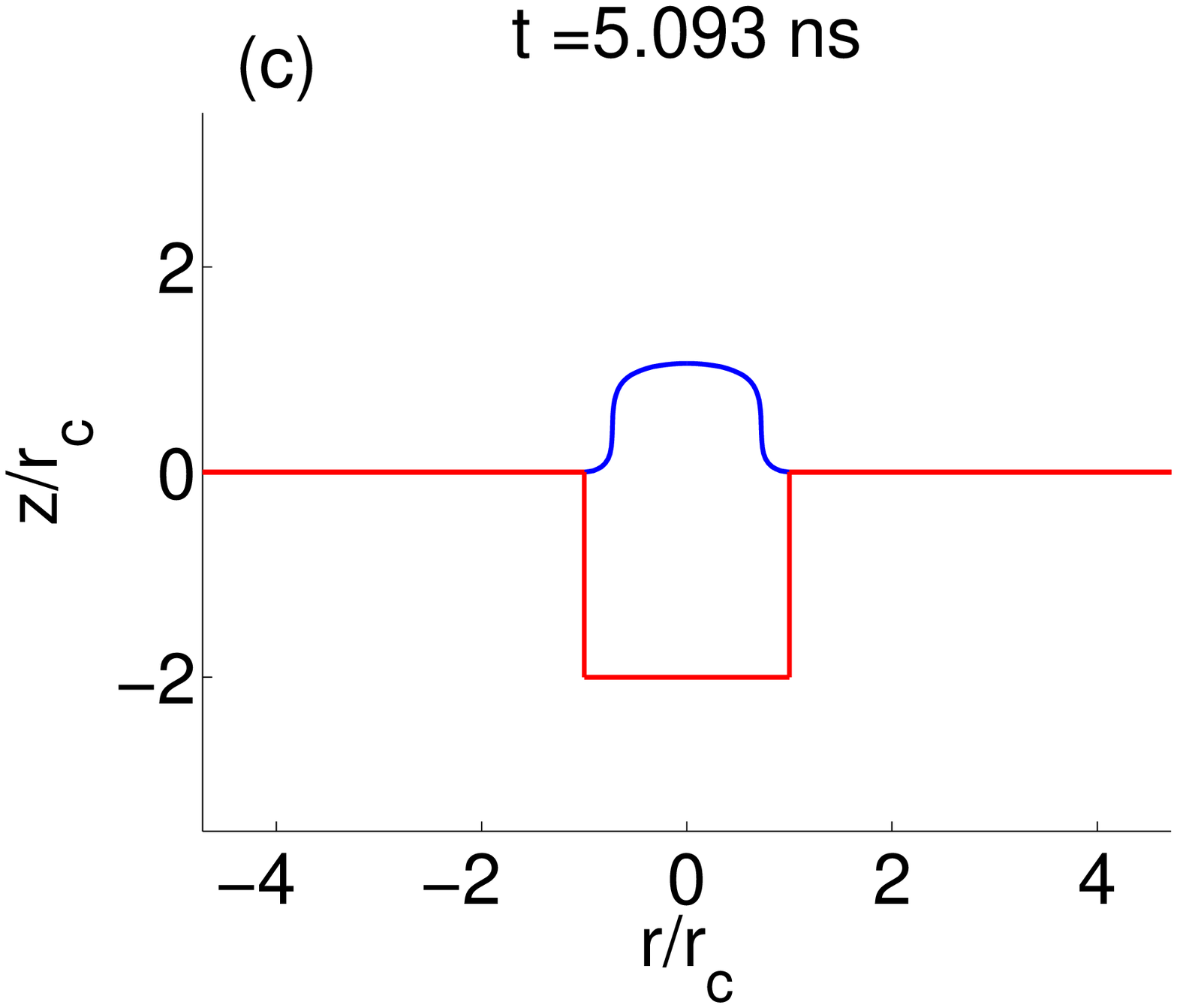}
\includegraphics[width=0.45\columnwidth]{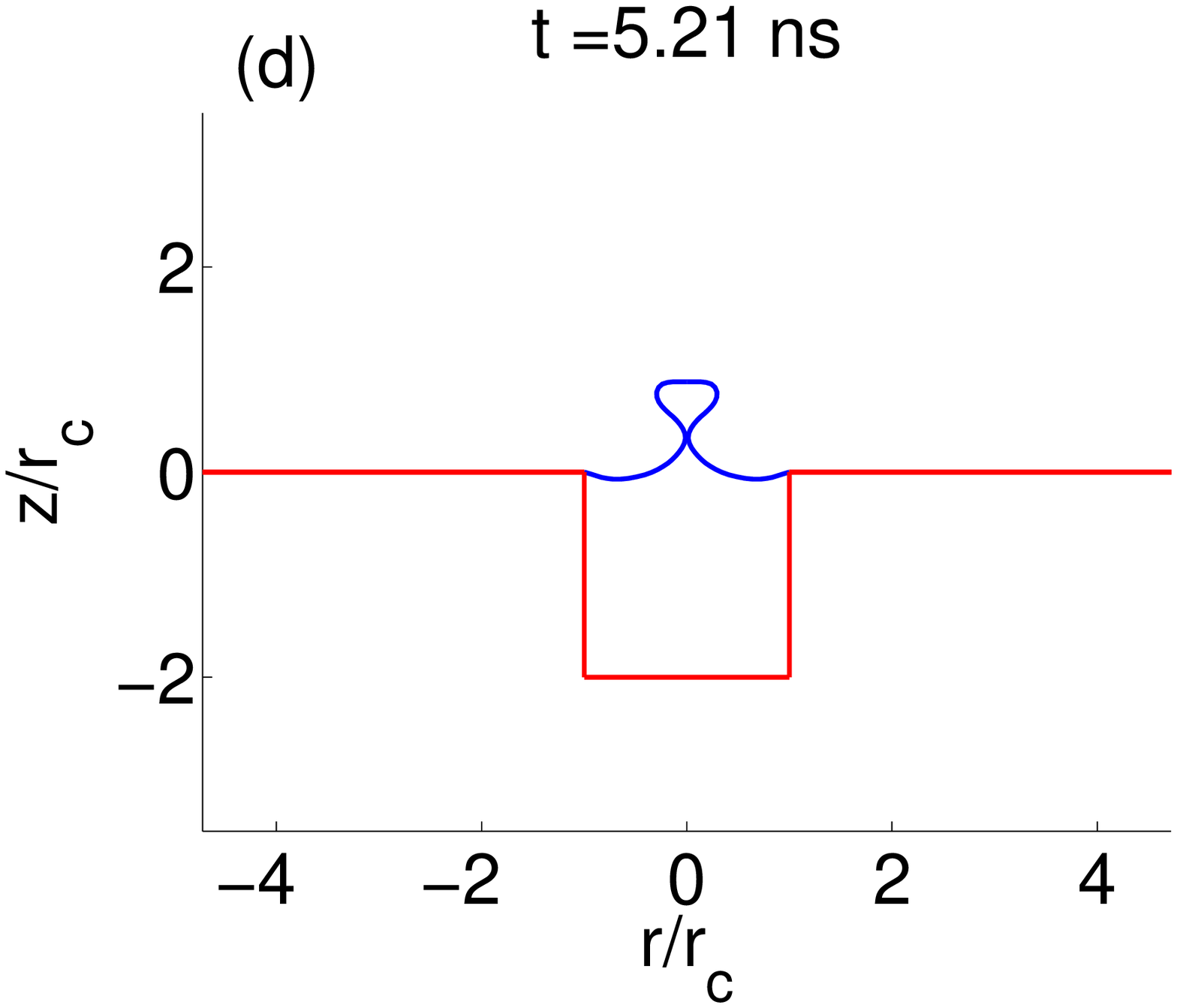}
\includegraphics[width=0.45\columnwidth]{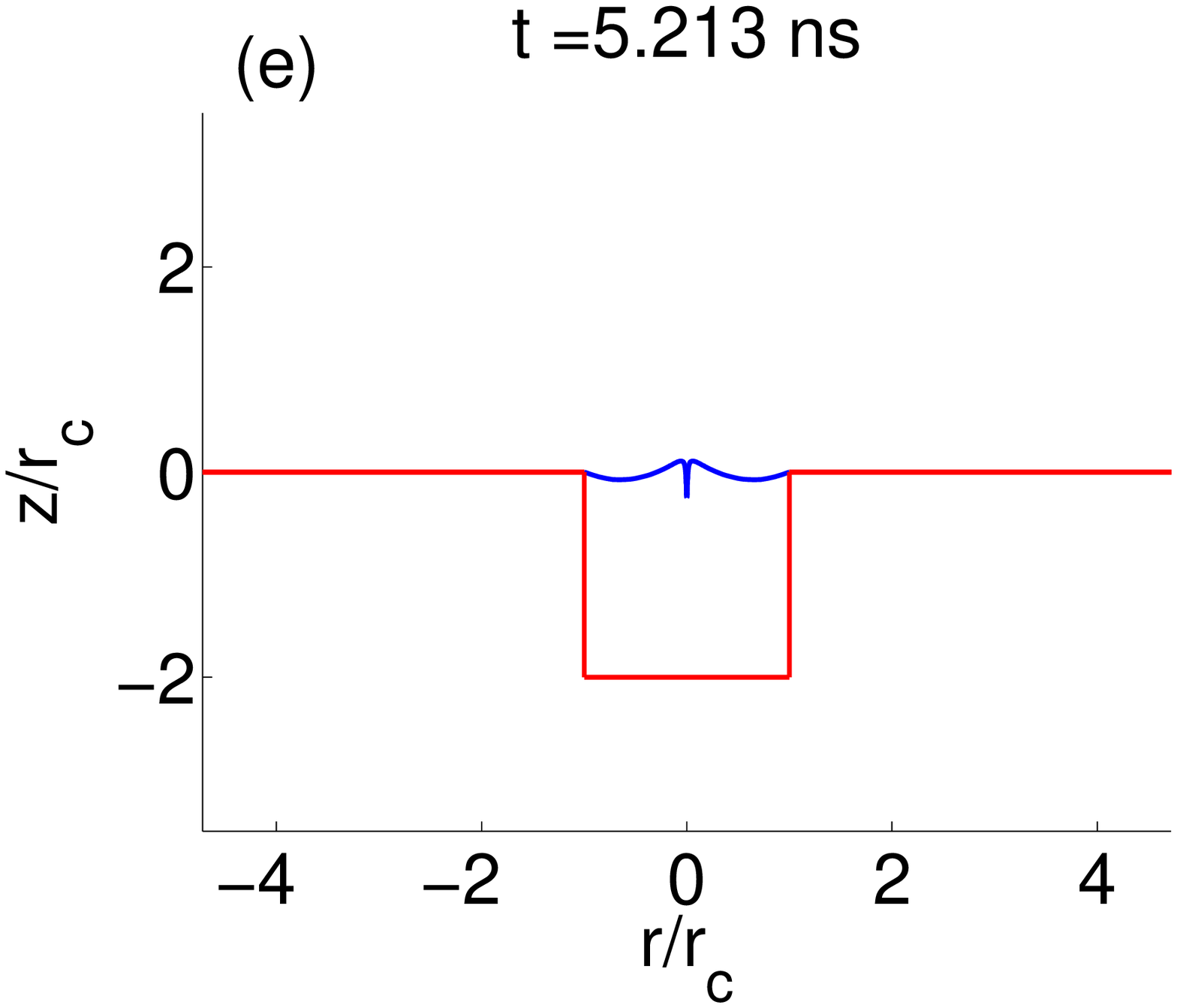}
\includegraphics[width=0.45\columnwidth]{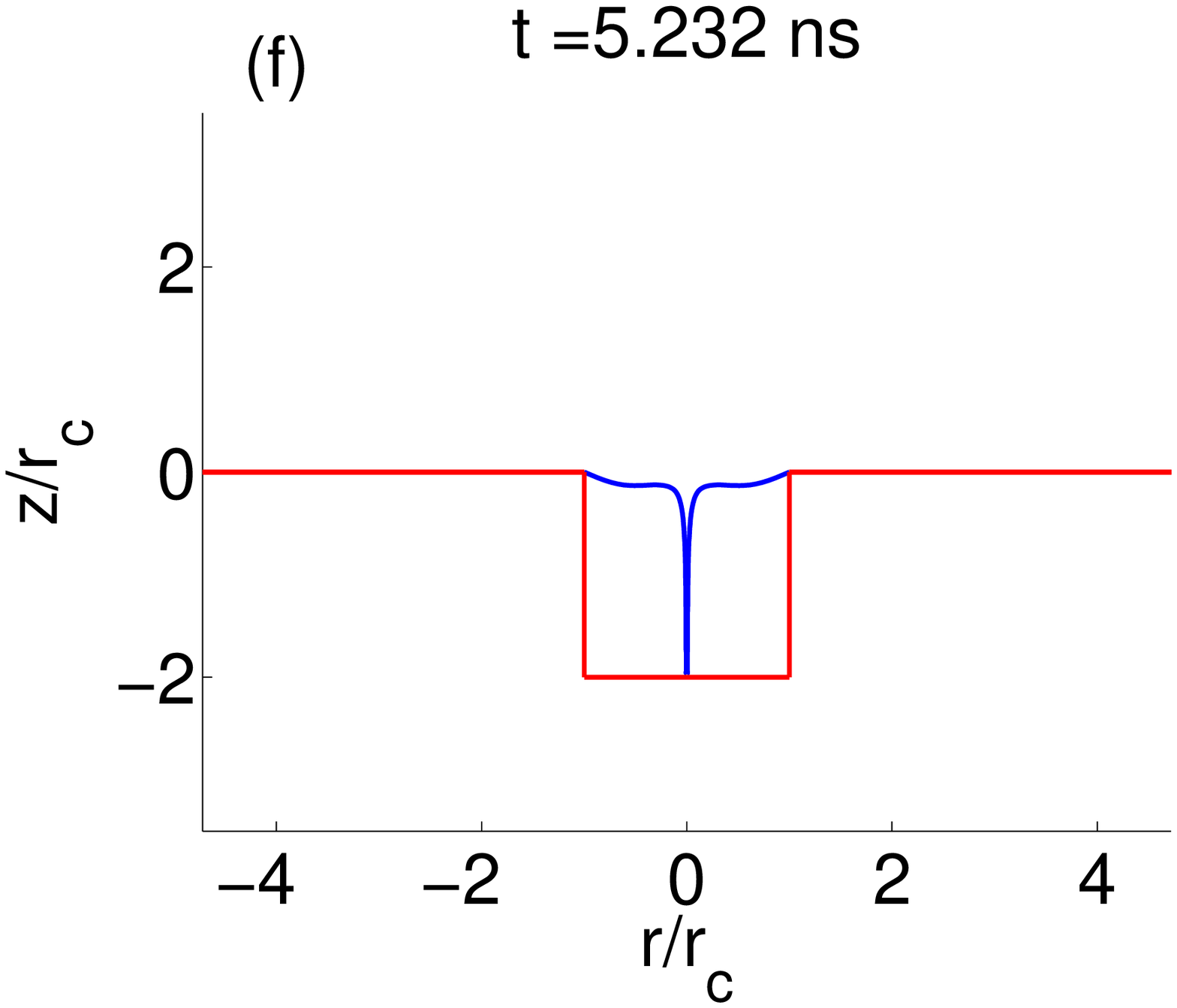}
\caption{(a) Initial configuration for a spherical bubble (blue) with radius $5r_c$ on top of the cavity (red). The (advancing) contact angle is $\theta=124^{\circ}$. Due to the low pressure inside the cavity the bubble starts to collapse (b) evolving into an almost cylindrical shape (c), which eventually closes on the axis of symmetry in a single point (d). From the pinch-off location a downward jet protrudes into the cavity (e) eventually hitting the cavity bottom (f). Here it would cause a splash filling the cavity with liquid.}\label{fig:simShape}
\end{center}
\end{figure}

 In order to shed light on this proposed deactivation mechanism, numerical simulations were carried out to elucidate the shape of the air-liquid interface during the bubble collapse. For this we used the boundary-integral method described in Ref.~[\onlinecite{bergmann09}] based on a potential flow description of the liquid dynamics.
The liquid-solid angle was prescribed to be $\theta_a=124^{\circ}$ which corresponds to the experimentally determined advancing contact angle of water on the substrates. At the starting point of the simulations the bubble was assumed to be a segment of a sphere with a radius significantly larger than the cavity radius $r_c$, see Fig.~\ref{fig:simShape} (a). The pressure inside the bubble was assumed to be uniform, satisfying the adiabatic relation $p_0/V_0^{1.4} = p_{bub}/V_{bub}^{1.4}$ with the initial pressure $p_0$ and $V_0$ the volume of the crevice, and $p_{bub}$ and $V_{bub}$ the instantaneous pressure and volume of the bubble, respectively. After release the bubble begins to shrink rapidly due to both surface tension and the low internal pressure. Eventually it evolves into an almost cylindrical shape as illustrated in Fig.~\ref{fig:simShape} (c). This air cylinder collapses radially and finally closes in a single point on the axis of symmetry, leaving a small air bubble entrapped above the pinch-off point, see Fig.~\ref{fig:simShape} (d). Towards pinch-off the liquid rushing radially inwards has to accelerate more and more to satisfy the requirement of mass conservation. When the advancing liquid front reaches the axis of symmetry a high pressure develops and the flow is deflected up and down to form two fast, needle-like water jets. The continuing collapse of the air cavity below and above the pinch-off point provides additional momentum to the two jets\cite{gekle09}. The downward jet protrudes deeply into the cavity until it hits the bottom of the cavity as illustrated in Fig.~\ref{fig:simShape} (e)-(f). For simplicity we neglected the upper bubble here which is expected to have only negligible influence on the downward jet. Upon reaching the bottom the impacting jet would form a violent, non-axisymmetric splash which cannot be captured by our numerical technique. Nevertheless, one can easily imagine the continuation of the process: as more and more liquid enters the cavity through the jet, the cavity is flooded with liquid, making a second nucleation impossible. We checked that the jet mechanism is present for pits in the size range studied here (50\,nm $<r_c<$500\,nm) and is independent of the initial bubble size, pit depth, and contact angle.

The previous explanation is not applicable to the case of strong mutual interaction between the bubbles, when the jet tends to be deflected away from the wall-normal~\cite{bremond06b}, although the nuclei are still observed to be de-activated after emission of the first bubble. However, during the expansion phase of the cavitation bubble (typically $ t_b \sim 10\mu \mathrm{s}$), gas diffuses from the cavity into the bubble, thus leaving the pit. The typical diffusion length scale is $\sim \sqrt{D t_b}=10 \mu$m, using the diffusion coefficient $D\sim10^{-5} \mathrm{m}^2/\mathrm{s}$. In the cases studied here the pits are $<1 \mu$m deep, i.e. much smaller than the typical diffusion length scale. Also, the volume of a typical bubble is $10^5 - 10^7$ larger than the volume of the nanopits, allowing the majority of the gas to move from the pit into the bubble.\footnote{Notice that this estimation does not take into account the confinement of the nanopits and their large aspect ratios (case D), which may lead to lower values of the diffusion coefficient.}
Unlike the wall-normal case the different jet dynamics does not return the gas to the pit, which therefore remains full of vapor and is easily filled by the liquid. These conclusions agree with our experimental results. In the cases A-D we observed no differences between strong (Fig.~\ref{fig_cav_focus}) and weak (Figs.~\ref{fig_exp_246},~\ref{fig_exp_495} \&~\ref{fig_exp_53}) bubble interaction: in both situations the pits were emptied after one nucleation event, in line with the explanations provided here.

This situation can be compared with the previously mentioned continuous and long-lived emission of bubbles in carbonated beverages and the related phenomena observed with enhanced surfaces frequently used in boiling heat transfer~\cite{messina81,qi04}. In none of these cases the bubbles collapse and therefore the primary mechanism for the filling up and consequent de-activation of the pits is present.

\subsection{Superhydrophobic nuclei}
\begin{figure}
\centering
        \includegraphics[height=2.5cm]{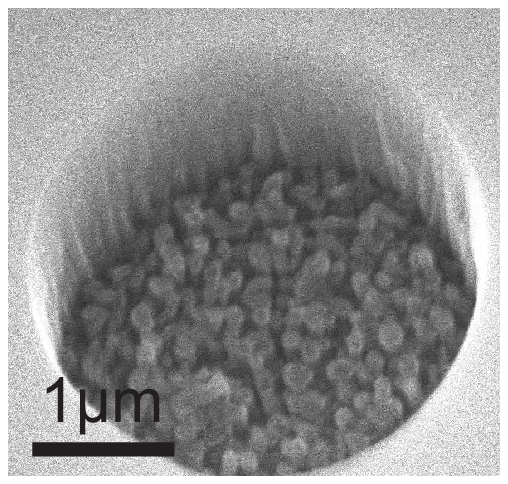}
        \includegraphics[height=2.5cm]{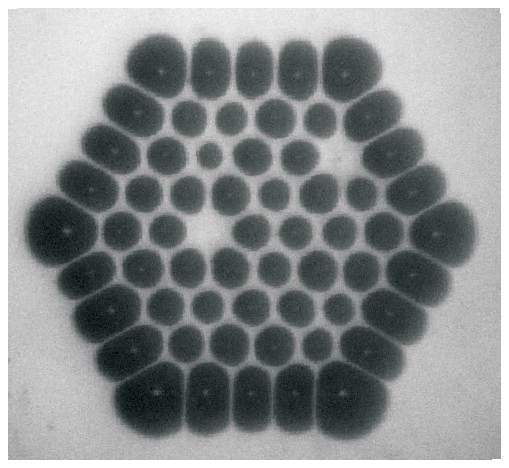}
        \includegraphics[height=2.5cm]{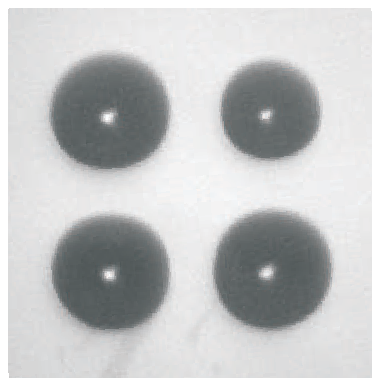}
        \caption{A superhydrophobic pit (left) can be nucleated hundreds of times, provided that the liquid jet in the bubble collapse phase is not directed into the pit. Center: a hexagonal pattern of superhydrophobic pits (100 $\mu$m in between the pits) after 230 nucleation events shows only 2 defects. Right: a square pattern (300 $\mu$m in between the pits) is completely intact after 100 shots.}
        \label{fig_sh}
\end{figure}

To show the importance of the liquid jet into the pits, experiments using superhydrophobic nuclei are illustrative. Cylindrical pits, etched in Si(100) with a diameter of 4 $\mu$m, were created with a superhydrophobic bottom layer (see Fig.~\ref{fig_sh}), consisting of hydrophobic pillars of $\sim 100$\,nm in diameter (created through a black silicon etching process~\cite{jansen95} and a hydrophobic fluoro-carbon top layer). The combination of hydrophobicity with roughness is known to create superhydrophobicity~\cite{quere02} with typical contact angles $> 160 ^{\rm{o}}$. When the bubbles were strongly interacting (leading to wall-parallel jets), we observed that the pits remained active nucleation sites even after hundreds of shots (see Fig.~\ref{fig_sh}). On the contrary, in the case of a single bubble (wall-normal jet) we observed that the micropits were deactivated after a few nucleation events. To explain this striking difference one really has to take the direction of the liquid jet into account. Apparently, the wall-parallel jet is not able to wet the superhydrophobic bottom of the pits, while the vigorous, ultra-thin jet directed towards the superhydrophobic bottom layer presumably pushes the liquid from the dewetted into the wetted state. From other work it is indeed known that a force may be required to overcome the energy barrier associated with this wetting transition~\cite{patankar04}. Once in the wetted state, the superhydrophobic pit is deactivated and cannot be nucleated again, apart of course from drying the whole sample.\\

\section{Conclusion}
In conclusion, Atchley \& Propseretti's 1989 crevice model of cavitation nuclei is experimentally verified using nanoscopic well-defined nuclei. Advanced etching techniques allowed us to create cylindrical pits down to 50\,nm in radius with high accuracy in both their size, depth and mutual position. Upon immersion in water, the hydrophobic nanopits trapped air and served as nucleation sites. Stepwise lowering of the acoustic minimum pressure allowed us to determine the nucleation threshold at which the pits start to cavitate. We found that the experimental results are in very good agreement with the theoretical predictions. This implies that in shock wave experiments the size of cavitation nuclei can be determined by measuring the pressure at which they start to nucleate.

Cavitation nuclei were deactivated after a single nucleation event, despite differences in width, depth and aspect ratios of the pits. The two mechanisms contributing to this effect are diffusion of gas out of the pit during the lifetime of the bubble and the subsequent aspherical collapse of the bubble. Numerical simulations show that in the case of weak bubble-bubble interaction, a sharp wall-normal liquid jet is formed which hits the bottom of the cavities, thus vigorously wetting the pits. Superhydrophobic nuclei can only be wetted thanks to this wall-normal jet. For strong bubble-bubble interactions with wall-parallel jets, superhydrophobic pits remain active nucleation sites, even after hundreds of nucleation events, in contrast to standard hydrophobic pits. In systems were one wants to control the number of cavitation nuclei which do not deactivate, superhydrophobic pits may find applications.

\section{Acknowledgements}
We are grateful to Clemens Padberg for the preparation of the samples with the nanopits and Johan Bomer for the samples with the superhydrophic pits. We acknowledge Holger Sch\"onherr for the AFM work involved. This work was supported by NanoNed, the nanotechnology program of the Dutch Ministry of Economic Affairs (Grant TMM.6413).

\end{document}